\documentclass{aa}  

\usepackage{graphicx}
\usepackage{txfonts}
\usepackage{lineno}
\usepackage{url}
\begin{document} 

   \title{ExPRES: a Tool to Simulate Exoplanetary and Planetary Radio Emissions}


   \author{C. K. Louis
          \inst{1,2,3}
          \and
          S. L. G. Hess
          \inst{4}
          \and
           B. Cecconi
          \inst{2,3}
          \and
          P. Zarka
          \inst{2,3}
          \and
          L. Lamy 
          \inst{2,3}
          \and
          S. Aicardi
          \inst{5}
          \and
          A. Loh 
          \inst{2}}

   \institute{Institut de Recherche en Astrophysique et Plan\'etologie (IRAP), CNRS, Universit\'e Paul Sabatier, Toulouse, France\\
            \email{corentin.louis@irap.omp.eu}
        \and
            LESIA, Observatoire de Paris, PSL Research University, CNRS, Sorbonne Universit\'es, UPMC Univ. Paris 06, Univ. Paris Diderot, Sorbonne Paris Cit\'e, Meudon, France
        \and
             USN, Observatoire de Paris, CNRS, PSL, UO/OSUC, Nan\c cay, France
        \and
            ONERA-The French Aerospace Lab, FR-31055 Toulouse, France
        \and
            DIO, Observatoire de Paris, PSL Research University, CNRS, Paris, France}

   \date{}

 
  \abstract
   {All magnetized planets are known to produce intense non--thermal radio emissions through a mechanism known as Cyclotron Maser Instability (CMI), requiring the presence of accelerated electrons generally arising from magnetospheric current systems. In return, radio emissions are a good probe of these current systems and acceleration processes. The CMI generates highly anisotropic emissions and leads to important visibility effects, which have to be taken into account when interpreting the data. Several studies showed that modeling the radio source anisotropic beaming pattern can reveal a wealth of physical information about the planetary or exoplanetary magnetospheres that produce these emissions.}
   {We present a numerical tool, called ExPRES (Exoplanetary and Planetary Radio Emission Simulator), which is able to reproduce the occurrence in time--frequency plane of CMI--generated radio emissions from planetary magnetospheres, exoplanets or star--planet interacting systems. Special attention is given to the computation of the radio emission beaming at and near its source.}
  {We explain what physical information about the system can be drawn from such radio observations, and how it is obtained. These information may include the location and dynamics of the radio sources, the type of current system leading to electron acceleration and their energy and, for exoplanetary systems, the magnetic field strength, the orbital period of the emitting body and the rotation period, tilt and offset of the planetary magnetic field. Most of these parameters can be remotely measured only via radio observations.}
   {The ExPRES code provides the proper framework of analysis and interpretation for past (Cassini, Voyager, Galileo\dots), current (\textit{Juno}, ground-based radiotelescopes) and future (BepiColombo, Juice) observations of planetary radio emissions, as well as for future detection of radio emissions from exoplanetary systems (or magnetic white dwarf--planet or white dwarf--brown dwarf systems). Our methodology can be easily adapted to simulate specific observations, once effective detection is achieved.}
   {}


\maketitle
%



\section{Introduction}
\subsection{Planetary radio emissions}

All planets in our solar system that possess an internal magnetic field are known to emit low frequency radio emissions, in wavelength domains extending from kilometer (below $\sim$100 kHz) up to decameter (a few tens of MHz --- in the case of Jupiter only). The frequency domain corresponds to the electron cyclotron frequencies ($f_{ce}$) close to the planet, revealing that the emission process is related to the electron gyration along the planet's magnetic field lines. Theoretical work and in situ observations of the terestrial, kronian and jovian radio sources permitted to elucidate the physical process at the origin of the radio emissions: the Cyclotron-Maser Instability (CMI), which occurs when an elliptically polarized wave resonates with the gyration motion of accelerated electrons \citep[see ][]{Wu85,Louarn,Zarka98,Treu06,Louarn2018,Lamy2018}. Under some circumstances --- notably a positive gradient of the perpendicular velocity distribution of the electrons --- the CMI mainly amplifies the wave on the extraordinary mode which can escape the source and propagate in free space as a radio wave.

The interest for planetary low-frequency radio emissions is driven by their relation with accelerated electrons. Those are also responsible for auroral emissions on top of the planet's atmosphere (over a broad spectral domains extending from Infrared to X-rays) and reveal the presence of field-aligned currents coupling the magnetosphere to the planet's ionosphere. Contrary to the other auroral emissions, radio emissions are not emitted on top of the planet's atmosphere but along a larger altitude range extending from the top of the ionosphere up to a few planet radii \citep[see review by][]{Zarka98}. The emission frequency is close to $f_\text{ce}$ in the source, itself proportional to the local magnetic field strength which decreases with increase altitude. Hence, the radio source altitude can be deduced from the frequency at which it emits. This property can be used to probe large altitude ranges above the auroras and to reveal, for example, the presence of acceleration regions \citep{Pot2007,Hess06,Hess2009b}.

\subsection{Visibility and variability of the sources}
\label{subsec:visibility}

The CMI is very sensitive to the plasma characteristics in the source, such as the density and temperature of the different electron populations, and the shape of the electron velocity distribution \citep{Pritchett84,Louarn96th}. These parameters not only condition and affect the amplification and the propagation  of the wave, but they also have a huge impact on the beaming pattern of the emission (see section \ref{sec:res}). Typical CMI emissions are radiated on a few degrees \citep[][]{Kaiser00, Zarka04b} of a given angle relative to the magnetic field. By symmetry around the magnetic field direction, the emission pattern is a thin hollow cone. This strong anisotropy of the radio emission beaming has two consequences regarding the observations of planetary auroral radio emission: (1) the observations have to be detrended from the source visibility effects before being interpreted --- no detection of the emission does not mean that no emission is produced ---, and (2) as the beaming pattern strongly depends on the plasma characteristics close to the source, the visibility of the emissions carries information about the plasma parameters at the source.

\begin{figure}
\centering
\includegraphics[width=\linewidth]{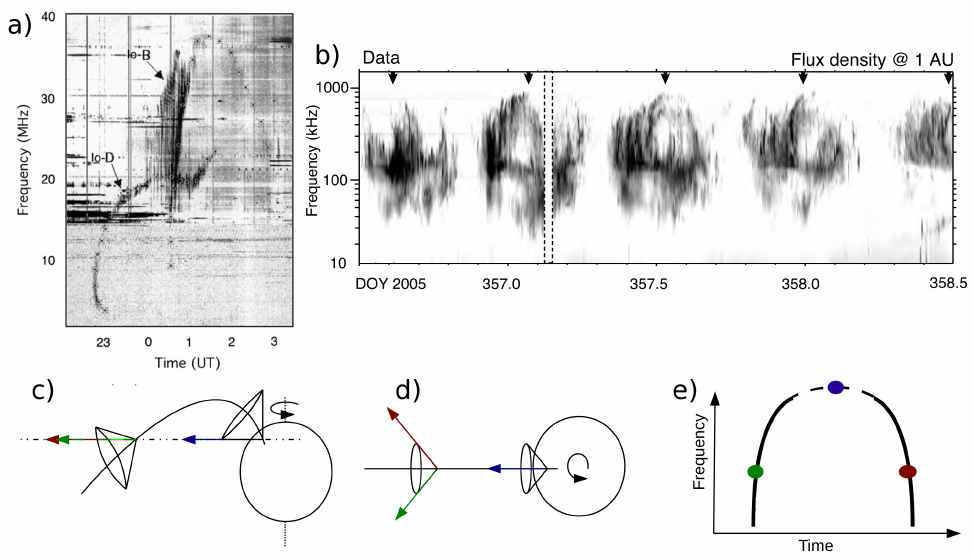}
\caption{a-b) Examples of time-frequency radio arcs. a) Radio arcs emitted by the Io-Jupiter interaction \citep{Quein98}, observed by Wind ($0$--$15$~MHz) and the Nan\c c Decameter Aarray ($15$--$40$~MHz). b) Radio arcs at Saturn, observed by Cassini, related to a sub-corotating hot spot in Saturn's magnetosphere  \citep{Lamy2008b}. c) and d) Side and polar views of the emission geometry, with two sources located along the same magnetic field line but at different altitudes. Arrows show the direction of propagation of the radio waves which can be seen by an observer located in the equatorial plane far from the planet. e) Dynamic spectrum of the emissions which would correspond to the above geometry.}
\label{fig_arc}
\end{figure}

Visibility effects are responsible for the ubiquitous arc shapes of the radio emission patterns in the time-frequency plane, as illustrated by the examples of jovian and kronian emissions displayed on Figure \ref{fig_arc}. Panels \ref{fig_arc}a and \ref{fig_arc}b respectively show typical emissions from Jupiter \citep{Quein98} and from Saturn \citep[including arcs generated by hot-spots in sub-corotation in Saturn's magnetosphere, ][]{Lamy2008b,Lamy2013}. The arc shape is a direct consequence of the hollow cone beaming pattern of the source, as shown on Panels  \ref{fig_arc}c--e in the simple case of a dipolar-axisymetric magnetic field. The radio emission from a radiating field line is received once when the source is on the western side of the observer's meridian (i.e., before the meridian relative to the sense of planetary rotation) and that one side of the cone is directed toward the observer. It is observed again when the field line is on the eastern side of the meridian and that the other side of the cone is directed toward the observer. As a result, for a fixed observer and a field line moving in the sense of rotation, radio emission at a given frequency can thus be observed twice, once or none depending on the geometry of the beaming pattern.

To interpret observations of planetary auroral radio emission, it is thus necessary to take into account their beaming pattern for inferring the effect of the visibility of the source. Some studies assumed constant beaming angles along the field lines \citep{Gen1985,Lec1998}, while others allowed the beaming angle to vary with the frequency \citep{Quein98,Hess2014}, in order to determine the source location. For some case studies, e.g. of specific radio arcs, a beaming pattern can be deduced from the observed arc shape assuming a source location (or possible source locations), as in \citet{Quein98}, or using the source location provided by an independent method such as radio goniopolarimetry \citep{Cecconi2009,Lamy2011}. Deriving physical source parameters directly form observations can be qualified of 'inverse modeling'. It allows to infer source and emission parameters that are consistent with a given observation of a radio arc, but it does not prove that only the observed arc should indeed be visible at the time of observation, i.e., it does not entirely de-trend the observation from the beaming effect. Moreover, the inverse modeling often implies to determine both the position and the beaming pattern from a single observation, whereas these two parameters are strongly coupled leading to degeneracy of the solution \citep{Hess2009b}. For a more global interpretation of radio dynamic spectra, a 'forward modeling' approach may be better: it consists of assuming source and emission parameters for computing a predicted dynamic spectrum, that is then compared to the observations. Matching of the predicted and observed dynamic spectra is a strong proof of the adequacy of the model to the reality of the source parameters and emission process. Of course, good matches may be obtained for non unique sets of parameters. Nevertheless, modeling of various observations corresponding to different viewing geometries is expected to remove this degeneracy and permit to better constrain the source conditions.

\subsection{Visibility modeling}

Implementing this forward modeling approach is the purpose of the numerical code described in the present paper, the Exoplanetary and Planetary Radio Emission Simulator (ExPRES). This code uses as inputs the geometry of the observation (observer and celestial bodies positions, sources location and magnetic field topology), the plasma parameters in the sources and their vicinity (density and temperature) as well as the characteristics of the wave-particle interaction generating the radio emissions constrained by the CMI theory. From these inputs, the code computes the beaming pattern of the radio sources, compares the direction of emissions to the direction of the observer and generates time-frequency visibility maps, that can be directly compared to observed dynamic spectra (Section \ref{sec:geo}).

In the following, we summarize the physics of planetary auroral radio emissions and how it is taken into account in ExPRES, starting with of the magnetospheric interactions powering the emissions and their impact on the CMI (Section \ref{sec:cur}). We then more specifically discuss the radio emission beaming (Section \ref{sec:theta}). We show that the different cases of auroral radio emission observed in our solar system can be described using a small number of parameters, which define both the location of the sources and their beaming pattern. Finally we describe the outputs parameters of the code, show some simulations examples and list simulations results already obtained with ExPRES (Section \ref{sec:conclusions}).

\section{ExPRES modeling}\label{sec:geo}

\begin{figure}[b]
\centering
\includegraphics[width=\linewidth]{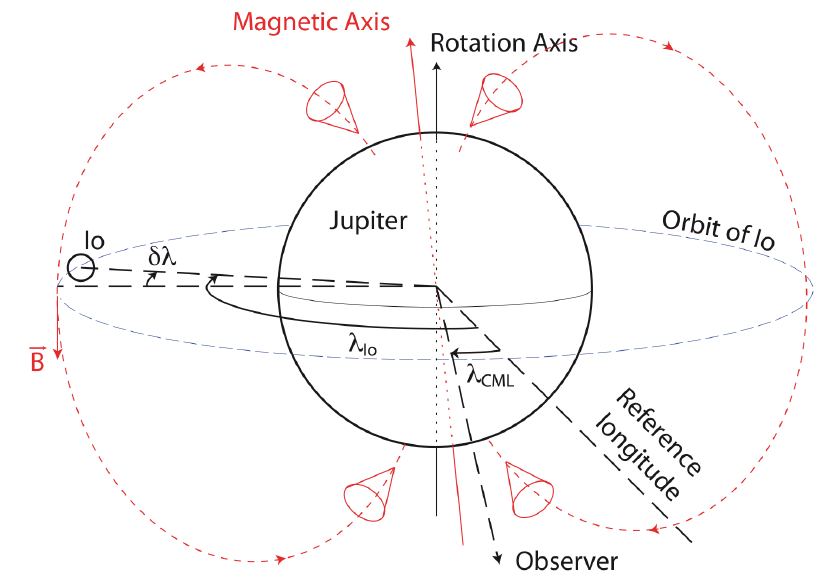}
\caption{Examples of the geometry involved in the modeling of the Io-controlled emissions of Jupiter. The emissions depends on the relative position of Io, Jupiter and the observer, in particular on the jovigraphic longitude of the observer ($\lambda_{CML}$ -- Central Meridian Longitude) and that of the active field line which can differ from that of Io $\lambda_{Io}$ by a lead angle $\delta\lambda$.}
\label{fig_io}
\end{figure}

The computation of a synthetic dynamic spectrum with ExPRES mostly relies on the observation geometry. An example is shown on Figure \ref{fig_io}, which sketches the geometry of the observation of emissions triggered by the Io-Jupiter interaction. For a given source-observer geometry (relative positions of the source and the observer at a given time and for a given emission frequency), magnetic field orientation in the source (depending on a magnetic field model), and beaming pattern (depending on electron densities and energies at the source, as discussed in Section \ref{sec:theta}), the code compares the direction of beamed RX mode waves with the source-to-observer direction. If the angle between these directions is smaller than the beam width, defined by the user, the corresponding time-frequency pixel in the synthetic dynamic spectrum is incremented by 1, otherwise it is not. By repeating this computation at all time-frequency steps of interest, for all elementary point sources constituting the user-defined radio source, a visibility map is generated in the time-frequency plane. ExPRES counts by default a standard intensity value of 1 (unit-less) for each radio source. More physical intensities can nonetheless be used to achieve realistic simulations \citep[see example in][]{Lamy2013}.

The set of parameters that the user needs to feed ExPRES with in order to perform a simulation run is:
\begin{itemize}
\item{The definition of the celestial bodies involved in the simulation (star, planet, satellite), with their respective positions at each time step (pre--computed or deduced from initial positions and orbital parameters), their rotation~/~revolution period, and their size and mass. These parameters permit to define completely the geometry of the celestial bodies involved in the simulation.}
\item{The position of the observer relative to the reference body at each time step.}
\item{The magnetic field and density models attached to the bodies of interest (planet, Sun, planet-star or planet-satellite system, \dots). The magnetic field model may be dipolar or of higher order, and the magnetic field lines need to be pre--computed. The density profiles may be that of an ionosphere (exponential decrease with distance from the center of the body), a stellar corona (decrease in inverse square of the distance from the center), a plasma disk (exponential decreases from both the equatorial plane and the body's center, with different scale heights), or a plasma torus (exponential decrease from the center of a torus of given radius around the body). These parameters enter in the definition of the characteristics of the radio emission, in particular its beaming angle (see section \ref{sec:theta}).}
\item{The spatial distribution of sources, i.e, the position of the magnetic field lines carrying the current (from their footprint longitude/latitude, their equatorial longitude/distance or their attachment to a satellite) and the frequency range over which the simulation is to be performed.}
\item{The information used to calculate the opening $\theta$ of the beaming cone (the distribution function and the energy of the resonating electrons involved in the radio emission, see Section \ref{sec:cur}) and the thickness $\Delta \theta$ of the beaming conical sheet.}
\end{itemize}

While the source-observer geometry and the magnetic field, plasma density and energy models at the sources are user--defined in ExPRES, the beaming angle $\theta$ is self--consistently computed by the code from beaming models described in Section \ref{sec:theta} that are based on the CMI theory described in section \ref{sec:res}. Nevertheless, test simulations may be performed with a predefined beaming angle profile versus frequency (e.g. $\theta(f)$ constant or linearly varying with the frequency).


ExPRES simulations can be adjusted to observations by varying one or several input parameters (e.g., the electrons energy), and the quality (and possibly uniqueness) of the obtained fit permits to derive the value of the corresponding parameter(s) \citep{Hess2009b}. 

\section{Radiosource locations and characteristics}\label{sec:cur} 
\subsection{Magnetospheric currents}

The auroral radio emissions simulated by ExPRES are deeply related to the dynamics of the magnetosphere of magnetized planets. The magnetosphere is the region where the plasma motion is dominated by the magnetic field of the planet. In first approximation, the plasma and the magnetic field are frozen--in. However, the magnetosphere undergoes constraints --- both external (solar wind flow) and internal (centrifugal forces, satellite outgassing\dots) --- that force the plasma motion relative to the magnetic field. This motion creates an electric field with an associated current, which in turn induces a magnetic field that distorts the magnetic field lines to try to keep them frozen in the plasma. Although this is a crude and over-simplified way to summarize magnetospheric physics, it is as a first approximation sufficient to understand the physics involved in the present paper.

The electric currents follow the magnetic field lines --- because the conductivity parallel to the magnetic field is far larger than that perpendicular to it --- and close in or above the planet's ionosphere. At some point along the magnetic field line, usually within $\sim1$ planetary radius above the ionosphere, the increase in magnetic field strength generates a resistance to the current and thus an electric field parallel to the magnetic field. Electrons accelerated by these electric fields precipitate in the planet's atmosphere and generate auroras. Outside the accelerating region, these electrons move adiabatically, at least in first approximation, so that their parallel velocity $v_\|$ decreases with the magnetic field strength $B$:
\begin{equation}
v_\|^2=v^2-\mu B
\end{equation}
with $\mu$ the magnetic moment of the electrons. If $v_\|$ goes to zero before the electrons reach the ionosphere (where they get lost by collisions and power Infrared to Ultraviolet auroras), they are reflected  and can participate to the radio emission generation via the CMI.

The three main magnetospheric interactions that lead to electron acceleration are:
\begin{itemize}
\item{The flow of the solar wind, which deforms the magnetosphere to give it a comet-like shape. This leads to the convection of magnetic field lines from the front to the tail of the magnetosphere, with the generation of one or two convection cells, or to viscous interaction along the magnetospheric border. Along with convection cells, an auroral oval fixed in local time (although modulated by the planet rotation) is formed at the footprints of field lines returning toward the front of the magnetosphere \citep{Dungey1961}. Viscous interactions rather lead to less structured and less stationnary auroras, with a strong local time asymmetry \citep{Axford1961,Delamere2010}.}
\item{The centrifugal motion of plasma generated inside the magnetosphere. As it moves outward, the conservation of the momentum forces the plasma azimuthal velocity to decrease, in which case it does not corotate anymore with the magnetic field. A current is then generated which re-accelerates the plasma and enforces corotation. This interaction leads to a very stable auroral oval which is fixed in longitude \citep{Cowley2001}.}
\item{The interaction of the planetary magnetic field with satellites. When the latter are deeply embedded within the magnetosphere of their parent planet, the plasma that surrounds them (e.g. their ionosphere) is forced to deviate from corotation with the planet's magnetic field. This also generates currents \citep{Neu80,Saur04}.}
\end{itemize}

To model this large diversity of interactions and radio auroral counterparts, ExPRES offers several possibilities in the choice of the magnetic field lines which are carrying the radio sources. One can model a full auroral oval, or only part of it (i.e. an auroral arc), either fixed in local time, fixed in longitude (corotating), or in sub--corotation. The position of this oval is defined by a fixed magnetic latitude (i.e. the distance of the field line at the equator). This permits to model auroral radio emissions resulting from solar wind--magnetosphere interactions or from the centrifugal motion of plasma in the magnetosphere, as well as 'hot-spots' related to sub--corotating regions of the magnetosphere \citep[such a those observed at Saturn and modeled using ExPRES in][]{Lamy2008b}. Simulations performed in \citet{Hess_exo} showed the typical morphology of the radio emissions in each of these cases.

ExPRES also permits to impose the active (radio-emitting) field line to be fixed in the frame of a satellite (with a possible longitude shift between the satellite and the active magnetic filed line in order to model the propagation time of the current perturbation between the satellite and the planet), thus allowing to simulate satellite--planet interactions \citep{HessGRL,Hess2009b, Louis2017b, Louis2017c, Louis2017a}. This option may also be used to simulate the interaction between a star and an exoplanet \citep{Hess_exo}, or interactions between magnetized stars \citep[][with a model similar to ExPRES]{Kuz2012}.

\subsection{Electron acceleration}

Besides the distribution of radio sources along specific magnetic field lines, one needs to define the characteristics of the current system associated to electron acceleration, because they will determine the density and temperature of the electrons inside the radio sources, as well as their distribution function. These characteristics will in turn constrain the beaming pattern of the radio source (see following sections).

The currents created by the interactions described in the previous section are of two types: stationary and transient currents. In these currents, the electrons move in an adiabatic way (in first approximation), thus their parallel velocity $v_\|$ (respectively perpendicular $v_\perp$) decreases (or increases) with intensity $B$ of the magnetic field. If  $v_\|$ reaches zero before the electrons reach the atmosphere, they reach their mirror point and are then reflected.

For stationary current systems, the magnetic mirroring of high latitude electrons acts as a resistive effect and generates an electric potential gradient along the magnetic field lines. This gradient is usually localized and takes the form of one or several double layers (i.e., discrete potential drops), between the ionosphere and a few radii above it. Between these double layers, the electron density is lower than along the magnetic field lines around (which carry no current) thus forming auroral cavities. In such cavities, the background cold plasma is absent and the electrons are accelerated downwards by potential drops. Thus theses accelerated electrons have a horseshoe distribution, resulting from a parallel acceleration followed by pitch angle increase due to the adiabatic motion of the electrons in an increasing magnetic field. Note that at Jupiter, the very low electron density at high latitude does not require the presence of such a cavity. However, the electrons are accelerated to maintain the continuity of the current.

The information about the onset of the interaction propagates along the magnetic field lines at the Alfv\'en velocity. A transient current system is generated during a time corresponding to at least the travel time at the Alfv\'en velocity between the interaction site (e.g. the equator in the case of a satellite--magnetosphere interaction) and the planetary ionosphere. Because of the large size of the current system (several planetary radii in this example), this transient phase can last for a long time and may even be longer than the interaction time itself \citep[see e.g.][for te Io--Jupiter case]{Neu80,Gur1981,Saur04}.

In this case, electron acceleration is due to the parallel electric field associated to kinetic Alfv\'en waves above the planet's ionosphere. This electric field is modulated at the Alfv\'en wave frequency and does not form electric potential drops, and hence does not form directly auroral cavities either \citep[although cavities may slowly build up due to the excitation of ion acoustic waves, ][]{Hess2009b,Matsuda2012}. In this case, the electron density in the current system is the same as outside of it, with the cold component of the plasma remaining present, and the electrons distribution is either a ring or a Kappa-like distribution \citep{Swift,Hess07,HessIFT}. 

\subsection{Unstable electron distributions}\label{sec:res}

Wave--particle resonance is reached when the Doppler-shifted pulsation of the wave in the electron's frame ($\omega - k_\|v_{r_\|}$) is equal to that of the gyration motion of resonant electrons ($\omega_c\Gamma_r^{-1}$). $k_\|$ and $v_{r_\|}$ are respectively the parallel components (to the magnetic field lines) of the wave vector and of the velocity of the resonant particle. $\omega_c=eB/m_e$ is the electron cyclotron pulsation, with $B$ the magnetic field amplitude and $e$ and $m_e$ the electron's charge and mass. $\Gamma_r$ is the Lorentz factor associated with the resonant electron motion.

The resonance condition thus writes:
\begin{equation}
\omega = \Omega_{c_r}+k_\|v_{r_\|}\label{eqn:res0}
\end{equation}
with
\begin{equation}
\Omega_{c_r} =\omega_c\Gamma_r^{-1}=\omega_c\sqrt{1-v_r^2/c^2}
\end{equation}

In the weakly relativistic case ($v_r<<c$), the resonance equation can be rewritten as:
\begin{equation}
v_{r _\perp}^2 + \left(v_{r_\|} -\frac{k_\| c^2}{\omega_c}\right)^2 = c^2 \left(\frac{k_\|^2 c^2}{\omega_c^2}+2 (1-\frac{\omega}{\omega_c})\right)
\end{equation}

Thus the resonance equation is that of a circle in the $[v_\|, v_\perp]$ velocity space, whose center $v_0$, located on the $v_\|$ axis, is given by:
\begin{equation}
v_0=\frac{k_\| c^2}{\omega_c}
\end{equation}
which can be rewritten as:
\begin{equation}
v_0=\frac{\omega}{\omega_c} cN\cos\theta\label{eqn:theta}
\end{equation}
where $\theta$ is the direction of the emission ($k cos \theta = \vec{\textbf{k}}\vec{\textbf{b}} = k_\|$) and $N=ck/\omega$ is the refractive index value \citep[which must be taken into account to reproduce the observations, as shown by][]{Ray08}. 

The resonance condition can thus be rewritten as:
\begin{equation}
\omega=\Omega_{c_r}+k_\|v_{r_\|}=\Omega_{c_r}\left(1-\frac{v_{r_\|}}{c}N \cos\theta\right)^{-1}\label{eqn:res}\\
\end{equation}

Therefore, the resonance condition does not only define the pulsation of the amplified wave $\omega$, but also the direction of the emission $\theta$.

The resonance equation is under--constrained as there are two variables ($\omega$ and $\theta$) to determine for a single equation. To solve it one must consider another constrain brought by the CMI amplification equation, which states that the amplification occurs for positive gradients of the electron perpendicular velocity distribution around the resonant velocities $v_r$ \citep{Wu85}. Given an electron distribution, it is possible to determine which sphere has the maximum positive gradients along its border and then to determine the angle of emission $\theta$. Then the resonance equation gives the emission frequency $f$. Subsequently, we will indifferently call 'frequency' the frequency $f$ or the pulsation $\omega$ , since $\omega= 2\pi f$.

\begin{figure}
\centering
\includegraphics[width=\linewidth]{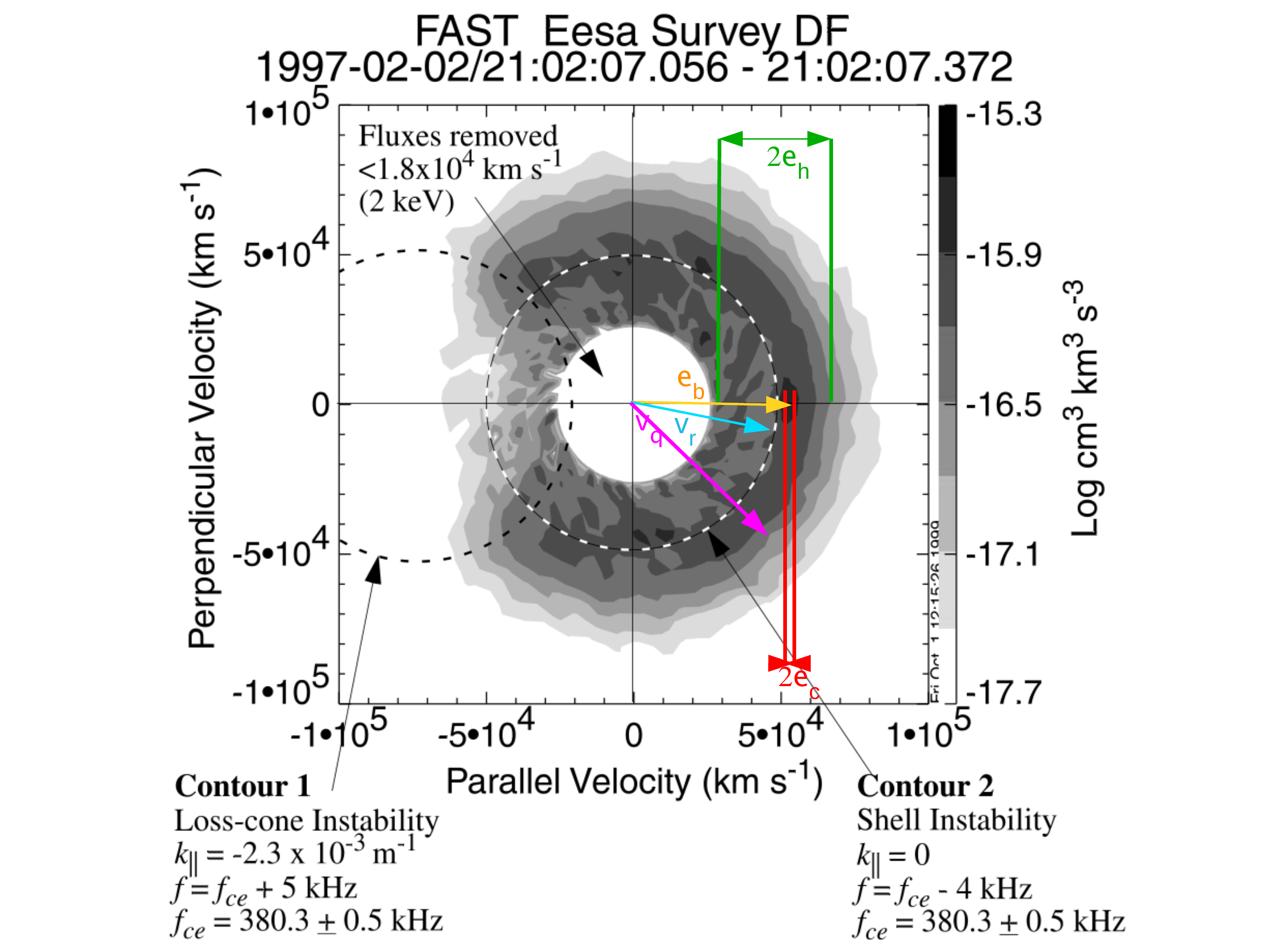}
\caption{An unstable electron distribution measured by FAST in the Earth auroral region \citep{Ergun2000}. This kind of distribution is typical in the auroral regions emitting radio emissions, it consists in a cold (a few eV) Gaussian distribution with a warmer (a few 100's~eV)  halo which is shifted in energy by a few keV and which pitch angle is scattered due to magnetic mirroring. We assumed a distribution function of that kind in our modeling of the beaming angle from auroral cavities. Velocity dispersion corresponding to the cold and halo temperature ($\epsilon_c$ and $\epsilon_h$) and the velocity shift corresponding to the beam energy ($\epsilon_b$) are shown. The resonant electron velocity ($v_r$) and the equivalent "thermal" velocity ($v_\theta$) are shown too.}
\label{fig_dist}
\end{figure}

Unstable electron distributions are common in the auroral regions. They may be so-called horseshoe distributions as that shown on Figure \ref{fig_dist}, or ring distributions which are incomplete horseshoes with a limited pitch angle spread of the electron velocity \citep{Su08}. These distributions may fulfill various resonance conditions, the two main ones being the oblique wave resonance (corresponding to $v_0\ne0$ and generically called loss-cone-driven) and the perpendicular wave resonance (corresponding to $v_0=0$ and generically called 'shell-driven') \citep{Wu85,Hess07}. The oblique mode resonance circle lies inside the loss-cone of the electron distribution and is tangent to it, where the distribution gradient generates the largest amplification rate, whereas the perpendicular mode resonance circle is tangent to the inner edge of the shell. These two resonance circles are shown by dashed lines on Figure \ref{fig_dist}.

 These modes differ by the frequency of the emission, deduced from the resonance equation (Eq.\ \ref{eqn:res}). The perpendicular emission frequency is obtained using $v_{r_\|}=0$ and is always smaller than the cold electron cyclotron frequency:
\begin{linenomath*}\begin{equation}
\omega_{shell}=\Omega_{c_r}<\omega_c\label{eqn:f_shell}
\end{equation}\end{linenomath*}

The emission frequency of the oblique mode is obtained from the center of the circle tangent to the loss-cone boundary at a point that can be obtained through elementary trigonometry and Eqs.\ \ref{eqn:theta} and \ref{eqn:res} \citep{Wu85,HessGRL}:
\begin{linenomath*}\begin{equation}
v_0=v_r/\cos\alpha\Rightarrow\omega_{lc}\simeq\omega_c\Gamma_r>\omega_c\label{eqn:lc}
\end{equation}\end{linenomath*}
where $\alpha$ is the resonant electrons pitch angle (i.e. the loss-cone angle), which depends on frequency:
\begin{linenomath*}\begin{equation}
\cos\alpha=v_{r_\|}/v_r=(1-\omega_c/\omega_\mathrm{c_{max}})^{-1/2}
\end{equation}\end{linenomath*}
with  $\omega_\mathrm{c_{max}}$ the electron cyclotron frequency at the altitude below which the electrons are lost by collision with the planetary atmosphere. 

Contrarily to perpendicular emission, oblique emission is emitted above the cold plasma electron cyclotron frequency. The difference in frequency between these modes has important consequences due to the characteristics of the dispersion relation of the right--handed waves in the plasma \citep{Lassen26}. 

While the bulk electron velocity is zero in the plasma rest frame, the velocity of individual electron is not zero. One thus needs to introduce in the refraction index expression a Lorentz factor $\Gamma_\text{th}$, translating the non--zero velocity of the electrons in the plasma rest frame \citep{Pritchett84,Louarn96th}. Exact computation of this relativistic correction may be complicated \citep{Pritchett84}, but it can be estimated by introducing an equivalent 'thermal' velocity $v_\text{th}$, so that the mean energy of the electrons in the plasma rest frame is $m_ev_\text{th}^2/2$ \citep{Louarn96th,Mottez10}. This relativistic term is of importance, as it participates --- along with the ratio between the plasma and the electron cyclotron frequency ($\omega_p/\omega_c$) --- in the determination of the mode(s) on which the waves can be emitted (or not).

Thus, in a non--zero temperature plasma, the expression of the refraction index is:
\begin{linenomath*}\begin{eqnarray}
N^2&=&1-\frac{\omega_p^2}{\omega^2-\frac{\Omega_{c_\text{th}}^2\sin^2\theta}{2(1-\omega_p^2/\omega^2)}\left(1+\sqrt{1+\left(\frac{2\omega(1-\frac{\omega_p^2}{\omega^2})\cos\theta}{\Omega_{c_\text{th}}\sin^2\theta}\right)^2}\right)}\label{eqn:AH}\\
\Omega_{c_\text{th}}&=&\omega_c\Gamma_\text{th}^{-1}=\omega_c\sqrt{1-v_\text{th}^2/c^2} \label{eqn:Omegath}
\end{eqnarray}\end{linenomath*}

\begin{figure}
\centering
\includegraphics[width=0.8\linewidth]{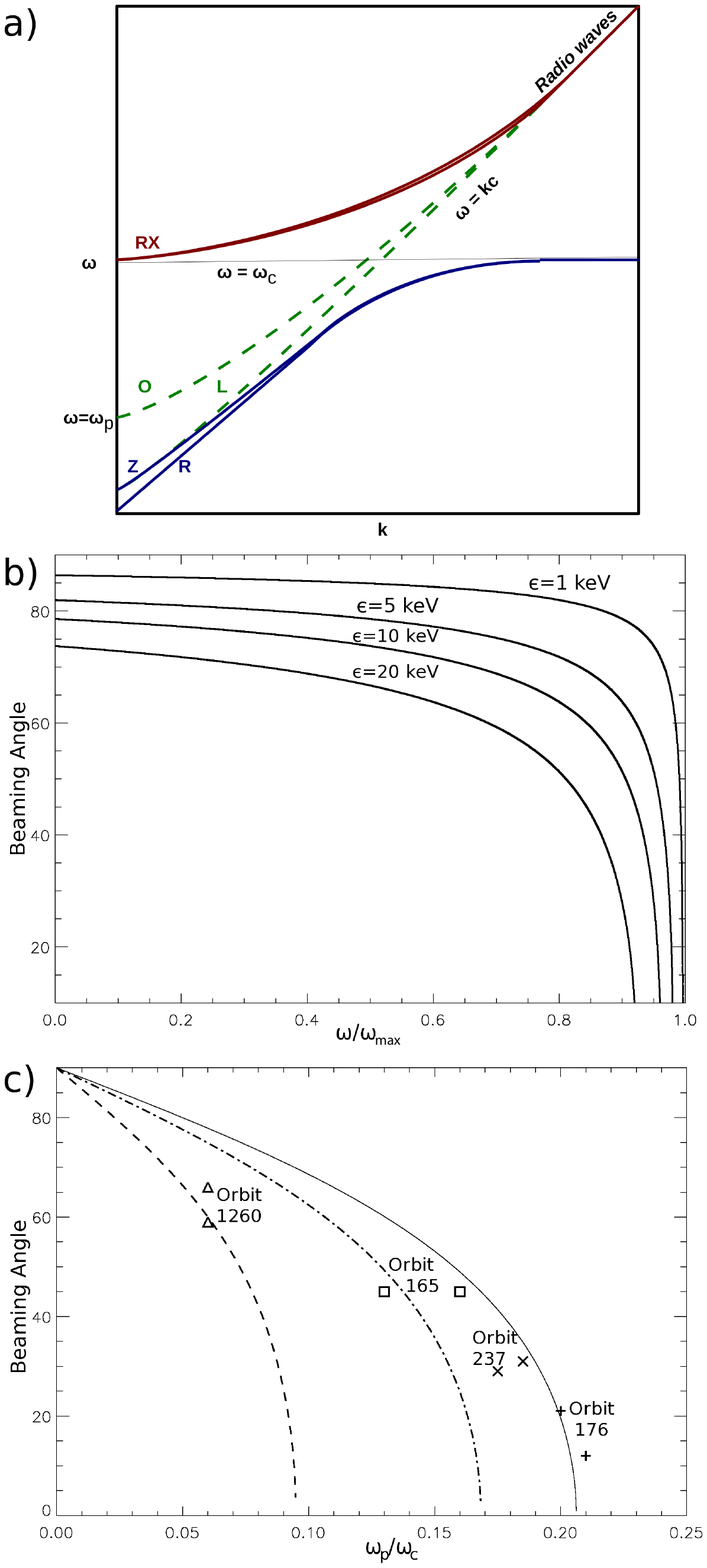}
\caption{a) Sketch of the dispersion relation of the electromagnetic waves in a plasma, following the Appleton-Hartree equation. Only R-X and L-O modes can escape the plasma and become radio waves. b) Beaming angle of the oblique emissions as a function of the ratio of the emission frequency to the surface electron cyclotron frequency, for different values of the electrons energy and $N=1$. c) Beaming angle of the perpendicular emissions as a function of $\omega_p/\omega_c$ outside of a cavity. Symbols correspond to AKR beaming angle measurements by \citet{Louarn96obs}. Lines are theoretical beaming angles computed using Eqs. \ref{eqn:stn} \& \ref{eqn:sba}, constrained by density measurements of \citet{Louarn96obs} for orbit 176 (solid line), orbit 1260 (dashed) and orbit 165 (dot-dashed).}
\label{fig_beam}
\end{figure}

From Equation \ref{eqn:AH} one sees that a cutoff frequency (N=0) exists for Right--Handed (RH) modes. Below this frequency, the wave is on the Right--Extraordinary R--Z mode, and above it, the wave is on the Right--Extraordinary R--X mode. Only the R--X mode connects to the $\omega=ck$ dispersion relation of freely propagating radio waves, whereas the R--Z mode is trapped inside its source region (see Figure \ref{fig_beam}a). Hence, R--X mode waves are the only RH waves observed as radio emissions \citep{Louarn96obs}. The 'choice' of the CMI between the R--Z and R--X modes is determined by two factors: the wave pulsation $\omega$, which is constrained by the resonant electron velocity, and the cutoff frequency, which depends on $\omega_p/\omega_c$ and $v_\text{th}$. In the auroral regions the cutoff frequency is always within a few percent of $\omega_c$. Lower values of $\omega_p/\omega_c$ and/or higher values of $v_\theta$ lead to smaller values of the cutoff frequency --- that can even lie below $\omega_c$ ---, whereas higher $\omega_p/\omega_c$ and/or lower $v_\text{th}$ lead to larger cutoff frequencies.

ExPRES computes the cutoff frequency at each point of the user--defined radio source from Eq. \ref{eqn:AH}, and the wave frequency from the electron velocity distribution (see section \ref{sec:res}), and considers that emission is produced only if the wave frequency is above the cutoff frequency. Thus ExPRES needs to be fed with the local magnetic field strength (that defines $\omega_c$), the local plasma density (that defines $\omega_p$), and the mean energy of the electrons in the source (that defines $v_\text{th}$).

Note that the modes discussed above all have pure circular polarization. Elliptically polarized waves amplified by the CMI can split in L--O and R--X modes on index gradients \citep{Melrose80,Shaposhnikov,Louarn96obs}. However, in the following we will only consider R--X emissions, because they are the dominant ones in most auroral emissions observed, and for that reason they are the only ones simulated by the ExPRES code so far.

\section{Radiation pattern}\label{sec:theta}
\subsection{Transient (Alfv\'enic) current system}

In the case of a transient current system, electrons are accelerated by Alfv\'en waves electric fields and a ring \citep{Hess07} or a Kappa-like \citep{Swift} distribution forms. Also, the Alfv\'en waves do not generate a deep auroral cavity devoid of cold plasma \citep{Mottez10}. In these conditions, the oblique R-X mode is the one that will be favoured. Its beaming angle is obtained by solving together equations \ref{eqn:AH}, \ref{eqn:theta} and \ref{eqn:lc}. First, the refraction index is obtained by equating  the value of the center of the resonance sphere $v_0$ in Eqs. \ref{eqn:theta} and \ref{eqn:lc}:
\begin{linenomath*}\begin{equation}
N=\frac{\Gamma_r^{-1}}{\cos\alpha}\frac{v_r}{c}\frac{1}{\cos\theta} =\frac{\chi}{\cos\theta}
\end{equation}\end{linenomath*}

The dispersion relation is obtained from the dielectric tensor \citep{Stix62} and can be written as:
\begin{linenomath*}\begin{equation}
AN^4-BN^2+C=A \chi^4-B \chi^2\cos^2\Theta+C\cos^4\Theta=0\label{eqn:abc}
\end{equation}\end{linenomath*}

Setting $\nu_p=\omega_p^2/\omega_c^2$ and using the resonance equation (Eq. \ref{eqn:res}) and the notation of \citet{Stix62}:
\begin{linenomath*}\begin{equation}
S=1-\frac{2\nu_pc^2}{v_r^2+v_{th}^2} ; P=1-\frac{2\nu_pc^2}{2c^2+v_r^2} ; D=\nu_p\frac{2c^2-v_r^2}{v_r^2+v_{th}^2}
\end{equation}\end{linenomath*}
the coefficients of Eq. \ref{eqn:abc} can be rewritten to obtain a second order equation in $\cos^2\theta$:
\begin{linenomath*}\begin{eqnarray}
S\chi^4+\left[\chi^4(P-S)-\chi^2(PS+S^2-D^2)\right]\cos^2\theta& &\nonumber\\
\ +\left[P(S^2-D^2)-\chi^2(PS-S^2+D^2)\right]\cos^4\theta& &\\
=a\cos^4\theta+b\cos^2\theta+c=0& &\nonumber
\end{eqnarray}\end{linenomath*}
The solution for the R-X mode is:
\begin{linenomath*}\begin{equation}
\cos^2\theta=\frac{-b+\sqrt{b^2-4ac}}{2a}
\end{equation}\end{linenomath*}

For a transient current system, the beaming angle inside the source computed by ExPRES is the exact solution of the above equation. To perform the calculation, ExPRES needs the user to specify the accelerated electron mean energy ($v_r^2$) and the plasma temperature ($v_\text{th}^2$), whereas the ratio between plasma frequency and electron cyclotron frequency is deduced from the density profiles associated with the celestial bodies considered (see section \ref{sec:geo}).

Panel b of Figure\ref{fig_beam} shows the evolution of the beaming angle as a function of the ratio between the emission frequency $\omega$ and the maximal cyclotron frequency reachable $\omega_\mathrm{c_{max}}$ for different resonant electron energies $\epsilon$. The plasma frequency was considered to be negligible, so that $N=1$ in that case.

\subsection{Steady-state current system}
\subsubsection{Refraction in the source}

At the source, the beaming of the radio emissions generated by a steady-state current system is more simple than in the previous case: if the R--X mode can be amplified, waves are beamed perpendicular to the magnetic field. However the emission is most of the time observed to be generated inside of a plasma cavity and to be refracted on its border \citep{Louarn96obs,Louarn96th}. This lead to a very complex situation due to the fact that the beaming angle outside the cavity depends on the cavity profile. There is a wide variety of possible cavity profiles and very few constraints on them. Moreover, depending on the shape of the cavity, waves can be partially trapped and resonate inside the cavity, with an impact on the radio beaming. As a consequence, it is not possible to define an exact general solution to the problem of the beaming angle of a radio source inside a cavity. Even for a well--defined cavity profile, computation of the beaming requires a ray-tracing algorithm, which needs too much computational power to be integrated in ExPRES.

ExPRES thus uses an approximate solution, assuming that the refraction mainly occurs inside the cavity and not on its borders. The need for refraction inside the source was already emphasized by \citet{Louarn96th}, and its existence is related to the presence of a gradient of refraction index inside the cavity due to the gradient of magnetic field strength. This gradient is mainly parallel to the magnetic field. Thus, the major difference between ExPRES computation and reality is that ExPRES assumes that the wave reaches a region where the refraction index is $N=1$ inside the source, in which case refraction on the cavity border has a low impact on the final beaming angle, whereas in reality a slight increase of the refraction index in the cavity allows the wave to escape it, so that a large part of the refraction takes place on the cavity border. Note that the assumption made in ExPRES is likely to be a good approximation in the case of a large cavity (in terms of its perpendicular size relative to the radio wavelength), because in that case a significant fraction of the refraction is indeed expected to take place inside the cavity. In such a large cavity, strong trapping of the wave is also unlikely, and thus no effect is expected on the radio beaming. It is also well suited for cases in which hot plasma is observed out of cavities (generally at very high latitude) such as in the case of the Saturn SKR source crossing \citep{Lamy2010,Lamy2011}.

Under the above assumption, assuming that the index gradient is parallel to the magnetic field and using Snell--Descartes law, computation of the radio beaming angle simply becomes:
\begin{linenomath*}\begin{equation}
\sin\theta=N\label{eqn:stn}
\end{equation}\end{linenomath*}
with the refraction index $N$ being that of a pure X mode (purely perpendicular to the magnetic field) with $\omega=\Omega_{c_r}$, i.e.:
\begin{linenomath*}\begin{equation}
N^2=1-\frac{\omega_{p_{source}}^2(1-\frac{\omega_{p_{source}}^2}{\Omega_{c_r}^2})}{\Omega_{c_r}^2-\Omega_{c_\text{th}}^2-\omega_{p_{source}}^2}\label{eqn:nsh}
\end{equation}\end{linenomath*}
It appears clearly from this equation that in order to have a real refraction index (permitting wave propagation), the electron mean energy (i.e. 'temperature') must be larger than the resonant electron mean energy so that $\Gamma_\text{th}>\Gamma_r$ and thus $\Omega_{c_r}^2-\Omega_{c_\text{th}}^2>0$.

\subsubsection{Electron distribution model}

Equation \ref{eqn:nsh} is very sensitive to the variation of each parameter, and thus the relation between them has to be set very carefully from physical considerations relevant to the source considered. This is particularly the case for the electron distribution.

ExPRES models the unstable distribution as a bi--Maxwellian distribution at rest accelerated to a given energy. Thus, we define three characteristic electron energies: the beam energy ($\epsilon_b$) which corresponds to the energy of accelerated electrons (usually several keV), the electron's core thermal energy ($\epsilon_{c}$) which is the temperature of the core of the distribution of the background electrons (usually a few eV to a few tens of eV), and the electron's halo temperature ($\epsilon_{h}$) which is the mean energy of supra-thermal electrons (usually a few hundreds of eV). The electron density inside the source is taken proportional to that outside of the source (i.e., that of the plasma before acceleration), with a proportionality coefficient deduced from current conservation (density $\times$ velocity is constant). The plasma frequency in the source is thus deduced from the ratio between beam and core energies:
\begin{linenomath*}\begin{equation}
\omega_{p_{source}}^2=\sqrt{\frac{\epsilon_{c}}{\epsilon_b}}\omega_{p}^2=\eta\omega_{p}^2\label{eqn:wpsource}
\end{equation}\end{linenomath*}

The halo temperature is used to determine the resonant electron energy. This energy is not that of the beam, because the resonance circle of a shell driven CMI passes through the region of largest $\nabla_{v_{r_\perp}}f({\bf v}_r)$, i.e., within the inner edge of the shell and not along the peak of the distribution. For a shifted-Maxwellian distribution of the energies in the beam (with a standard-deviation $\epsilon_{h}$), the largest positive gradient is obtained for an energy $\epsilon_b-\epsilon_{h}$. Thus the difference between the electron's mean and resonant energies is  :
\begin{equation}
v_\text{th}^2-v_r^2=\frac{2\epsilon_{h}}{m_e} \label{eqn:vth}
\end{equation}

As noted in \citet{Mottez10}, using $\epsilon_{c}$ instead of $\epsilon_{h}$ would not permit the generation of an X--mode wave (the plasma needs to be 'hot'), hence our assumption of the presence of a halo, which is usually observed at energies of a few hundreds eV in magnetospheric plasmas \citep[cf.\ Figure \ref{fig_dist} and][]{Ergun2000}.

Depending on what the user knows as physical parameters, two choices are possible with ExPRES. Either the user specifies the plasma density in the source $\omega_{p_{source}}$, or it only gives the global magnetospheric density (not directly that of the sources).

In the first case (if the user knows $\omega_{p_{source}}$), using Equations \ref{eqn:Omegath} and \ref{eqn:vth}, the refraction index of Equation \ref{eqn:nsh}  becomes:
\begin{linenomath*}\begin{equation}
N^2=1+\frac{1-\frac{\omega_{p_{source}}^2}{\Omega_{c_r}^2}}{1-\frac{\omega_c^2}{\omega_{p_{source}}^2}\frac{2 \epsilon_h}{m_e c^2}}
\end{equation}\end{linenomath*}

In the second case (if the user does not know $\omega_{p_{source}}$), using Equation \ref{eqn:wpsource} the refraction index of Equation  \ref{eqn:nsh} then becomes:
\begin{linenomath*}\begin{equation}
N^2=1+\frac{1-\eta\nu_p \Gamma_r^2}{1-\frac{2\epsilon_{h}}{\eta\nu_pm_ec^2}}<1\ \text{(R--X mode)}\label{eqn:sba}
\end{equation}\end{linenomath*}

Figure \ref{fig_beam}-c displays the beaming angles computed using the above equation, with parameters $\eta$ and $\epsilon_{h}$ deduced from measurements in Terrestrial auroral cavities. $\eta$ was estimated from the densities inside ($\sim$1 cm$^{-3}$) and outside (deduced from measured $f_{pe}$) of the cavities  \citep{Louarn96obs}, and $\epsilon_{h}$ was taken $\simeq900$~eV, consistent with the distributions measured by \citet{Ergun2000}. With measured beam energies $\epsilon_b=3-10$~keV, the electrons' resonant energy is about $2-9$~keV. The modeled beaming curves are in good agreement with the observed values of the beaming angles (symbols) for the observations of Auroral Kilometric Radiation corresponding to the density measurements in \citep{Louarn96obs}.

\subsection{Refraction in the source vicinity}

The refraction index has an important effect inside the source since the resonance process occurs close to the X mode cutoff frequency, where the refraction index varies rapidly for small variation of the plasma parameters.

Outside of the source, the refraction index rapidly goes to 1, in which case the waves escape freely as radio waves, or to 0, in which case the waves meet a reflection layer and are reflected. This phenomenon may happen close to the source, where the local cyclotron frequency is still close to that inside the source (and thus close to the wave frequency), or far from it --- for example radio emissions from auroral region may be refracted in the equatorial plasma sheet. ExPRES only takes into account the former case, i.e. refraction in the source vicinity, because it does not include any ray--tracing algorithm.

This refraction effect outside the source differs from that inside the source by the fact that it is not symmetrical relative to the magnetic field vector. The ExPRES code models it under the approximation of planar refraction index iso--surfaces at each point of the wave propagation, with a refraction index varying only in the local meridian plane. Thus the gradient of refraction index is assumed to be null in the longitudinal direction, so that the normal to the refraction index planes has a null longitudinal component. The modification of the beaming angle is then obtained easily from the Snell--Descartes law.

\section{Simulation results, Conclusions and Perspectives}
\label{sec:conclusions}

\subsection{Versions of ExPRES}
\label{susbec:versions}

Each version of the code has seen new modules added and minor bugs fixed. Here is given a summary of the majors evolutions. 

The first versions (0.1, 0.2 and 0.3) of the ExPRES code were developed to predict and interpret the future observations of the NASA \textit{Juno} space mission to Jupiter. Indeed, unlike previous low frequency radio astronomy experiments \citep[such as Cassini/RPWS,][]{Gurnett2004}, which were able to measure the 4 Stokes parameters and the \textbf{k} vector of incoming waves \citep{Cecconi2009}, the radio experiment onboard  \textit{Juno} \citep{Matousek2007, 2017SSRv..213..219B}, named Waves \citep{Kurth2017},  only measures the total intensity of incoming radio waves versus time and frequency \citep[it also performs some limited direction--finding, but that proved effective for case studies][]{2017GeoRL..44.6508I}.  The development of the ExPRES tool was therefore necessary to determine the origin of the emissions.

Version 0.4 of the code allowed a generalization to Saturn and version 0.5 to exoplanets. For version 0.6, the code architecture was completely redesigned and the production of 3D movies was added. The current stable version is 1.0 and allows, from JSON formatted \citep{JSON} input files, to produce the simulation results in CDF \citep{CDF}.

\subsection{Output parameters}

Different output parameters can be selected. The default setup returns the following information at each time and frequency step:
\begin{itemize}
\item \texttt{Polarization}: polarization of the observed wave (if it is detected), making it possible to differentiate emission from the northern (<0) or the southern hemisphere (>0);
\item \texttt{Theta}: opening of the conical emission sheet;
\item \texttt{CML}: longitude of the observer;
\item \texttt{ObsLatitude}: planetocentric latitude of the observer;
\item \texttt{ObsDistance}: distance (in planetary radius) between the observer and the planet;
\item \texttt{SrcFreqMax}: maximum frequency on the active flux tube at each time step;
\item \texttt{Fp}: electronic plasma frequency in the sources;
\item \texttt{Fc}: electronic cyclotron frequency in the sources.
\end{itemize}
Additional information may be requested by the user, such as:
\begin{itemize}
\item the azimuth $\phi$ of the wave seen by the observer, i.e., the part of the conical emission sheet from which the wave seen by the observer comes from, with the reference $\phi = 180^\circ$ the direction of the wave towards the magnetic equator;
\item the equatorial longitude of the sources. For simulations of emissions induced by a moon, the longitude of the sources is the longitude of the moon;
\item the position (x, y, z) of the sources;
\item the local time of the observer (under development).
\end{itemize}

\subsection{Simulation results}

\begin{figure}
\centering
\includegraphics[width=\linewidth]{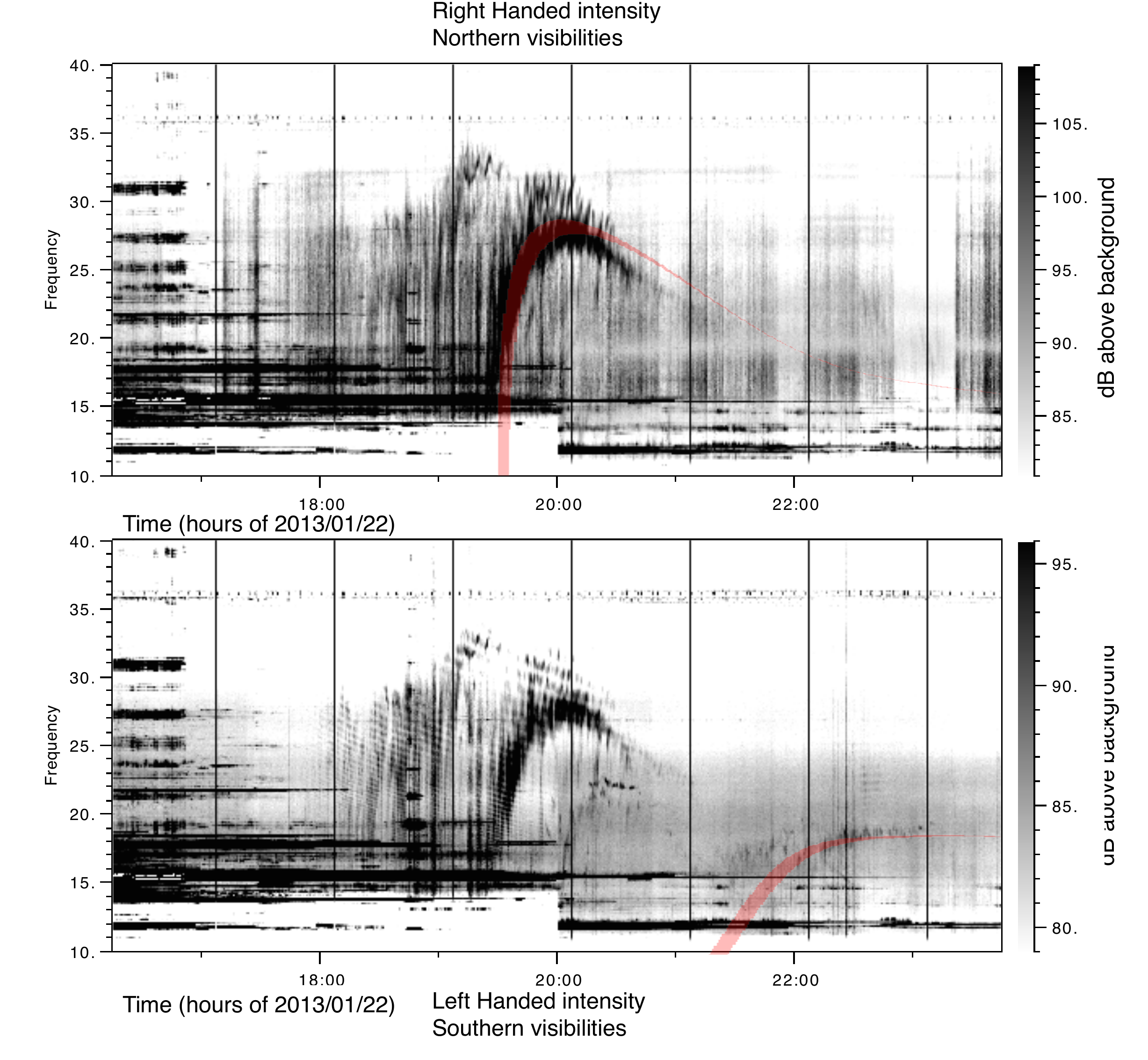}
\caption{Example of radio emissions (Io-controlled Jovian emissions) modeling using ExPRES overplotted on dynamic spectra observed by the Nan\c cay Decameter Array.The observation was performed on January 21st, 2013. Right-handed (top) and left-handed (bottom) polarization are separated. The observed emission (bright one in the northern hemisphere and dimmer one in the southern hemisphere) are well modeled by ExPRES (in red), assuming a source originating from the magnetic field line connected to the main Io Alfv\'en wing spot on Jupiter computed using the JRM$09$ model \citep{2018GeoRL..45.2590C}, and electrons with energies ($\epsilon_b$) of 2.5~keV (North) and 3.5~keV (South).} 
\label{fig_silfe}
\end{figure}

Figure \ref{fig_silfe} shows an observations with the Nan\c cay Decameter Array (NDA) of the Io--controlled Jovian emissions on January 21st, 2013. The two panels respectively correspond to the right--handed and left--handed emissions. This observation is dominated by an Io--B emission (emitted from a northern--dawn source, see Fig. \ref{fig_io}) that is mostly right--handed (consistent with its northern hemisphere origin), but a fainter Io--D left--handed--polarized emission (thus emitted from a southern--dawn source) is also visible later. These emissions can be modeled using ExPRES. The simulated dynamic spectra are overplotted  in red on top of the NDA data. The simulations matching best the observations are obtained assuming emissions from magnetic field line connected to the Io Main Alfv\'en Wing footprint, using the JRM$09$ magnetic field model \citep{2018GeoRL..45.2590C} and electrons with energies of 2.5~keV (in the northern hemisphere) and 3.5~keV (in the southern hemisphere). 

ExPRES has already been used to model observations, attempting to reproduce the time--frequency morphology of the radio emissions from Jupiter \citep{HessGRL, Hess2010_mag,Hess2009b,Ray08} and Saturn \citep{Lamy2008b,Lamy2011,Lamy2013}. It allowed these authors to determine the localization of the Io--Jupiter current circuit (downstream of the instantaneous Io--Jupiter field line) and to discover subcorotating field--aligned current systems at Saturn. These works also allowed them to put constraints on the energy of the accelerated electrons involved in the auroral radio components at these planets. Electrons energies of the order of a few keV were found at Jupiter and a few tens of keV at Saturn, consistent with UV observations.

ExPRES has also been used to predict and interpret radio emission from Jupiter, using a predefined dataset for the input parameters \citep[based on a parametric study, see][]{Louis2017c}. It allowed the authors to detect radio emissions induced by Europa and Ganymede \citep{Louis2017b} and interpret the time--frequency morphology of \textit{Juno}/Waves observations \citep{Louis2017a}, which was the initial goal of the code.

ExPRES was similarly used to simulate the radio environment of the Jupiter Icy Moon Explorer (JUICE) spacecraft planned to be sent to Jupiter by ESA, in order to estimate to what extent natural radio emissions from Jupiter's magnetosphere below 40 MHz would pollute the spacecraft radar measurements in the range 5--50 MHz \citep{CecconiRadar}.

Finally,  \citet{Hess_exo} studied the type of information that can be deduced, using ExPRES, from the morphology of the radio emissions --- still to be discovered --- from exoplanetary magnetospheres or star-planet systems involving plasma or magnetic interactions. These information include the magnetic- field strength and the rotation period of the emitting body (planet or star), the orbital period, the orbit's inclination, and the magnetic field tilt relative to the rotation axis or offset relative to the center of the planet. For most of these parameters, radio observations provide an unique mean of measuring them. 

\subsection{Accessibility}
The ExPRES code is written in IDL\footnote{Interactive Data Language, Exelis Visual Information Solutions, Boulder, Colorado}. The code is available under MIT licence on GitHub within the MASER\footnote{Measuring, Analyzing and Simulating Emissions in the Radio range} library repository: \url{https://github.com/maserlib/ExPRES}. The current version of the ExPRES code (V1.0) has been validated using IDL version 8.4. 

Precomputed ExPRES simulation runs are available through different interfaces described on the MASER project \citep{Cecconi_PV_2018,Cecconi_AGU2018} website\footnote{ExPRES page on the MASER website:  \url{http://maser.lesia.obspm.fr/tools-services-6/expres}}: (i) a web directory listing access; (ii) a virtual observatory access using the EPN-TAP \citep{Erard:2018kr}; and (iii) a streaming access using the das2 server framework \citep{piker_das2} through the MASER das2 Server. Table \ref{tab:access} provides URLs for all access interfaces.

\begin{table*}
{\scriptsize\begin{tabular}{llll}
Description&Interface&Access URL&Client\\
\hline
Web directory listing&HTTP&\url{http://maser.obspm.fr/data/serpe/}&Web Browser\\
Virtual Observatory dataset catalog&EPN-TAP&\url{http://voparis-tap-maser.obspm.fr:80/tableinfo/expres.epn_core}&VESPA or TAP client \citep{Erard:2018kr}\\
Streaming access&Das2&\url{http://voparis-maser-das.obspm.fr/das2/server}&Autplot \citep{Faden:2010jo}\\
\end{tabular}}
\caption{List of access interfaces for ExPRES precomputed dataset}\label{tab:access}
\end{table*}

The ExPRES code is also available for run-on-demand operations at \url{https://voparis-uws-maser.obspm.fr}. This computing interface requires an ExPRES JSON input configuration file. Examples of such configuration files are available through the web directory listing or virtual observatory catalog: each of the precomputed file is provided with its input configuration file. The JSON input files must conform to the corresponding JSON-scheme specification, the current version of which is available at \url{https://voparis-ns.obspm.fr/maser/expres/v1.0/schema#}.

\subsection{Perspectives}
ExPRES has proved to be a very useful tool for the modeling, analysis and interpretation of planetary radio emissions, the preparation and operation of planetary missions, as well as for the interpretation of the to--be--detected radio emissions from exoplanets. Such detections are expected soon as a result of extensive observing programs with giant radio telescopes such as LOFAR, UTR2, GMRT, NenuFAR and ultimately SKA.

Future uses will include the modeling of the time-frequency morphology of Jovian hectometer and broadband kilometer emissions \citep{BoischotJGR1981}, the modeling of the longitude--frequency morphology of Jovian hectometer and Io--independent decameter emissions \citep{Imai2008, Imai2011}, the search for induced radio emission of kronian satellite or the search for radio emission of Uranus and Neptune. Future developments may include a more quantitative treatment of the simulated emission intensity.

\section*{Acknowledgments}
Version 0.1 to 0.5 of ExPRES have been developed as SERPE ("Simulateur d'\'Emissions Radio Plan\'etaires et Exoplan\'etaires") by the LESIA laboratory under Observatoire de Paris and CNRS funding. Version 0.6.0 was developed by S.\ L.\ G.\ Hess on his own resources and with support from the LESIA, Observatoire de Paris for the online services. Later versions, including the current 1.0.0 version, were developed by the MASER team (\url{http://maser.lesia.obspm.fr}) which gathers personnel and funding from CNRS, CNES and Observatoire de Paris. Technical support from PADC (Paris Astronomical Data Center) is also acknowledged, with the very helpful contribution of P.\ Le Sidaner (DIO, Observatoire de Paris) and M.\ Servillat (LUTh, Observatoire de Paris). The Nan\c cay Decamater Array acknowledges the support from the Programme National de Plan\'etologie and the Programme National Soleil--Terre of CNRS--INSU. S.\ L.\ G.\ Hess thanks the MOP team of the LASP (Colorado Univ. Boulder) and R.\ Modolo (LATMOS/CNRS) for scientific support and fruitful comments. Part of the data distribution setup has been done in the frame of the Europlanet H2020 Research Infrastructure project, which has received funding from the European Union's Horizon 2020 research and innovation programme under grant agreement No 654208.

\bibliographystyle{aa}
\bibliography{ExPRES.bib}







\end{document}